\documentclass[3p,times]{elsarticle}

\usepackage{ecrc}

\volume{00}

\firstpage{1}

\journalname{}

\runauth{Jun-Sik Sin et al}

\jid{}

\jnltitlelogo{}

\usepackage{amssymb}
\usepackage[figuresright]{rotating}

\large
\begin{document}

\begin{frontmatter}



\dochead{}
%% e.g. \dochead{17th International Conference on Dynamical Processes in Excited States of Solids}

\title{Effect of orientational ordering of water dipoles on stratification of  counterions of different size in multicomponent electrolyte solution near charged surface - a mean field approach}
\author{\large Jun-Sik Sin}
\large
\ead{js.sin@ryongnamsan.edu.kp}
\author{\large Kwang-Il Kim}
\author{\large Chung-Sik Sin}
\address{\large Department of Physics, \textbf{Kim Il Sung} University, Taesong District, Pyongyang, DPR Korea}

\linespread{1.6}
\begin{abstract}
\large
We theoretically studied electric double layer by using the mean-field approach including the non-uniform size effect and the orientational ordering of water dipoles in electrolyte solution. Performing the minimization of the free energy, the resulting ion and water distribution functions are determined numerically in the process of solving the  Poisson's equation. The effect of non-uniform ionic size and orientational ordering of water dipoles on stratification of counterions of different size in multicomponent electrolyte solution near the charged surface are presented and discussed. Spatial dependence of relative permittivity and differential capacitance of electric double layer are also calculated for binary electrolyte solution and then compared with the predictions of the previous models. Effect of bulk concentration of counterions and ion size effects on surface charge density, excess portion of counterions, relative permittivity and differential capacitance are studied for multicomponent electrolyte solution in detail. 
\end{abstract}

\large
\end{frontmatter}

% \linenumbers

\section{Introduction}
\large
To advance the idea of electric double layer that von Helmholtz pioneered \cite%
{Helmholtz_annphys_1879}, Gouy \cite%
{Gouy_physfrance_1910} and Chapmann \cite%
{Chapman_philos_1913} proposed the original Poisson-Boltzmann approach, which does not consider the finite volumes of ions and neglects ionic correlations in electrolyte. The limitations of Poisson-Boltzmann approach yield unphysical ionic concentration profiles near highly charged interfaces and also inability to predict the overcharging phenomena.  

Furthermore, for electrolytes having counterions of different species, the Poisson-Boltzmann approach predicts that the concentration ratio of counterions of the same ionic valency is the same anywhere in the electric double layer, namely equal to that in bulk electrolyte. In particular, the approach predicts that for an equimolar mixture of divalent and univalent counterions the divalent ions should almost occupy the region of diffuse layer at a high surface charge. However, the experiments by Shapovalov and co-workers \cite%
{Shapovalov_jphyschem_2006, Shapovalov_jphyschem_2007} showed that near highly charged surfaces the predictions of Poisson-Boltzmann approach are no longer correct, but that instead the smaller ion (even in a mixture with divalent ones) can outcompetes the larger one, causing counterion stratification.

In order to overcome these limitations of Poisson-Boltzmann approach and account for ionic size effect, many researchers continuously developed realistic models of electric double layer. Stern \cite%
{Stern_zelektrochem_1924} modified the Poisson-Boltzmann approach considering the finite size effect of ions by combining the Helmholtz model with the Gouy-Chapmann model. Bikerman \cite%
{Bikerman_philmag_1942} tried to upgrade Boltzmann distribution functions  by taking into account the volume excluded by ions, but actually did not succeed to derive the new ion distribution functions and the corresponding Poisson equation. The role of  finite size of ions in electric double layer theory was firstly theoretically described by Wicke and Eigen \cite%
{Wicke_zelechem_1952}. In the last two decades, researchers considered finite volumes of ions and water molecules within statistical mechanics approach \cite%
{Borukhov_prl_1997, Borukhov_electrochimica_2000, Iglic_physfran_1996, Bohnic_electrochimica_2001, Bohnic_bioelechem_2002, Bohnic_bioelechem_2005}.  Recent theoretical studies \cite%
{Chu_biophys_2007, Tresset_pre_2008, Kornyshev_physchem_2007, Popovic_pre_2013, Li_pre_2011, Li_pre_2012, Boschitsch_jcomchem_2012} indicated that the voltage asymmetry of differential electric capacitance and stratification of counterions are strongly attributed to difference in sizes of ions.

Another main assumption of the conventional Poisson-Boltzmann approach is that regardless of electric field strength in an electrolyte, the relative permittivity of electrolyte is equal to the bulk value at any location near a charged surface. However, it is a well known fact that near a charged surface in an electrolyte the relative permittivity of the electrolyte solution varies according to the distance from the surface \cite%
{Onsager_amchem_1936, Kirkwood_chemphys_1939}. Although the Booth model \cite%
{Booth_chemphys_1951, Booth_chemphys_1955} has been widely used, the model does not account for sizes of both ions and water molecules in electrolyte solution. In \cite%
{Iglic_bioelechem_2010, Gongadze_bioelechem_2012, Gongadze_electrochimica_2013}, the authors developed a permittivity model based on orientational ordering of water dipoles, including size effects of ions and water molecules. The permittivity model well represents the fact that the permittivity of an electrolyte solution may be strongly decreased by orientational ordering of water dipoles and depletion of water molecules near a highly charged surface. However, all Poisson-Boltzmann approaches mentioned above are based on either both or one of following two assumptions: (1) ions of different species in electrolytes have an equal size, and (2) orientational ordering of solvent molecules is neglected. Recently, the  model presented in  \cite%
{Gongadze_bioelechem_2012} was  generalized by simultaneously taking into account non-uniform size of ions and orientational ordering of  water dipoles \cite%
{Gongadze_electrochimica_2015, Sin_electrochimica_2015}.

The approach in \cite%
{Sin_electrochimica_2015} is based on assumption of small volume shares of ions anywhere in the electrolyte solution, and uses implicit ion spatial distributions.

The authors of \cite%
{Gongadze_electrochimica_2015} generalized the model of \cite%
{Gongadze_bioelechem_2012} in order to consider the different size of positively and negatively charged ions in electric double layer, and obtained analytical expressions for ion spatial distribution functions. Mean-field approach in \cite%
{Gongadze_electrochimica_2015}  simultaneously takes  into account the asymmetry of ions of  binary electrolyte solution and  orientational ordering of water dipoles within lattice statistical mechanics approach \cite%
{Gongadze_bioelechem_2012}, where single negative and positively charged ions occupy more than one lattice site, while a  single water molecule  occupies just one lattice site.
	
In this paper we will develop an analytical Poisson-Boltzmann approach including both the different size effect of ions and water molecules and the orientational ordering of water dipoles. We will introduce a lattice statistics where more than one cell can be occupied by each ion as in \cite%
{Boschitsch_jcomchem_2012} and also by each water molecule for considering effects of different sizes of ions and water molecules. Unlike our previous approach \cite%
{Sin_electrochimica_2015}, the present approach yields the corresponding expressions for ion spatial distribution functions and is not based on the assumption of small volume shares of counterions.  In the section of results and discussion, considering or not orientational ordering of water dipoles, the role of the key factors \cite%
{Li_pre_2011, Li_pre_2012} for the stratification of counterions is estimated in comparison with previous studies. Next, the influence of different size of ions and water molecules and orientational ordering of water dipoles on spatial dependence of relative permittivity and on voltage dependence of differential capacitance are presented and discussed. Finally, the effect of the difference in sizes of counterions and  different bulk concentrations of counterions on surface charge density, excess portion of counterions, relative permittivity and differential capacitance are reported and explained. 

\section{Theory}
\large
We consider an electrolyte solution composed of different types of ions and water molecules in contact with a charged planar surface. The total free energy $F$ can be written in terms of the local electrostatic potential  $\psi\left(r\right)$ and the number densities of ions $n_{i}\left(r\right)$  $\left(i=1\ldots m\right) $  and water molecules 
$n_{w}\left(r\right)=\left<\rho\left(\omega, r \right)\right>_{\omega}$
 \begin{eqnarray}
	F=\int{d{\bf r}}\left(-\frac{\varepsilon_{0}\varepsilon E^2}{2}+e_{0}\psi\sum^{m}_{i=1}z_{i}n_{i}+\left<\rho\left(\omega\right)\gamma{p_{0}}E\cos\left(\omega\right)\right>_{\omega}-\sum^{m}_{i=1}\mu_{i}n_{i}-\left<\mu_{w}\left(\omega\right)\rho\left(\omega\right)\right>_{\omega}-Ts\right),
\label{eq:1}
\end{eqnarray}
where $\left<f\left(\omega\right)\right>_{\omega}={\int f\left(\omega\right)2\pi\sin\left(\omega\right)  d\omega}/\left(4 \pi\right)$ in which $\omega$ is the angle between the vector {\bf p} and the normal to the charged surface.  Here {\bf p} is the dipole moment of water molecules and {\bf E} is the electric field strength, while $z_{i}\left(i=1\ldots m\right)$ is the ionic valence of ions. The first term is the self energy of the electrostatic field, where $\varepsilon$ equals $n^2$ and $n=1.33$ is the refractive index of water. The second one represents the electrostatic energy of the ions in the electrolyte solution, where $e_{0}$ is the elementary charge. The third term corresponds to the electrostatic energy of water dipoles\cite%
{Gongadze_bioelechem_2012}, where  $\gamma=\left(2+n^2\right)/2$,  $p_{0}=\left|{\bf p}\right|$ and $E=\left|{\bf E}\right|$. The next three terms are responsible for coupling the system to a bulk reservoir, where $\mu_{i}$$\left(i=1\ldots m\right) $ are the chemical potentials of ions and $\mu_{w}\left(\omega\right)$ is the chemical potential of water dipoles with orientational angle $\omega$.
$T$ is the temperature and $s$ is the entropy density.

Consider a unit volume of the electrolyte solution near a charged surface. The entropy density is the logarithm of the number of translational and orientational arrangements of non-interacting   $n_{i}\left(i=1\cdots m\right)$ ions and  $\rho\left(\omega_{i}\right)\Delta\Omega_{i}\left(i=1 \cdots k\right)$ water molecules, where $\Delta\Omega_{i}=2\pi  \sin\left(\omega_{i}\right) \Delta\omega/\left(4\pi\right)$  is an element of a solid angle and $\Delta\omega=\pi/ k$. The ions and water molecule occupy volumes of $V_{i}\left(i=1\cdots m\right)$ and $V_{w}$, respectively.

Within a lattice statistics approach each particle in the solution occupies more than one cell of a lattice as in \cite%
 {Boschitsch_jcomchem_2012, Gongadze_electrochimica_2015}. Like in \cite%
 {Gongadze_electrochimica_2015}, orientational ordering of water dipoles as well as translational arrangements of ions is explicitly considered. 
Considering translational arrangements of ions and orientational ordering of water dipoles, the number of arrangements can be calculated as follows. As in \cite%
{Boschitsch_jcomchem_2012}, we first place $n_{i}$ ions of the volume $V_{i}$  in the lattice with unit volume. Accounting for the orientational ordering of water dipoles, we put in  $\rho\left(\omega_{i}\right) \left(i=0,1,...\right)$ water molecules of the volume $V_{w}$   in the lattice. The number of arrangements $W$ is written as 
 \begin{equation}
	W=\frac{N!}{\prod^{m}_{i=1}n_{i}!\cdot\lim_{k\rightarrow\infty}\prod^{k}_{i=1}\rho\left(\omega_{i}\right)\Delta\Omega_{i}!},
\label{eq:2}
\end{equation}
where
 \begin{equation}
  N\equiv\sum^{m}_{i=1}n_{i}+\lim_{k\rightarrow\infty}\sum^{k}_{i=1}\rho\left(\omega_{i}\right)\Delta\Omega_{i}!.
\label{eq:3}
\end{equation}
It should be noted that the above relation for the number of arrangements is exactly valid only for dilute solution or solution where ions and water molecules have the same size, while in the case of different sizes the relation is only approximate expression. 
 
 Expanding the logarithms of factorials using Stirling formula, we obtain the expression for the entropy density, $s=k_{B}\ln W$, 
 \begin{eqnarray}
	\frac{s}{k_{B}}=\ln W=N\ln N-\sum^{m}_{i=1}n_{i}\ln n_{i}-\lim_{k\rightarrow\infty}\sum^{k}_{i=1}\rho\left(\omega_{i}\right)\Delta\Omega_{i} \ln\left[\rho\left(\omega_{i}\right)\Delta\Omega_{i}\right],
\label{eq:4}
\end{eqnarray}
where $k_{B}$  is the Boltzmann constant.

Unlike in \cite%
{Sin_electrochimica_2015}, it is noted that the entropy density is naturally symmetric in different species of ions.

 All lattice cells should be occupied by either ions or water molecules \cite%
{Iglic_physfran_1996, Li_pre_2012, Li_pre_2011, Boschitsch_jcomchem_2012, Gongadze_bioelechem_2012, Gongadze_electrochimica_2015}
 \begin{eqnarray}
	1=\sum^{m}_{i=1}n_{i}V_{i}+n_{w}V_{w}.
\label{eq:5}
\end{eqnarray}

Using the method of undetermined multipliers, the Lagrangian of the electrolyte solution is 
 \begin{eqnarray}
	L=F-\int\lambda\left({\bf r}\right)\left(1-\sum^{m}_{i=1}n_{i}V_{i}-n_{w}V_{w}\right)d{\bf r},
\label{eq:6}
\end{eqnarray}
where $\lambda$ is a local Lagrange parameter.
The Euler$-$Lagrange equations are obtained and solved with respect to the functions $n_{i}$ and $\rho\left(\omega\right)$.
The variations of the Lagrangian with respect to $n_{i}$ and $\rho\left(\omega\right)$  yield equations from which we can get $n_{i}$ and $\rho\left(\omega\right)$  in the electrolyte solution:
 \begin{eqnarray}
	\frac{\delta L}{\delta n_{i}}=e_{0}z_{i}\psi-\mu_{i}+k_{B}T\ln\left(n_{i}/N\right)+V_{i}\lambda =0,   \left(i=1\cdots m\right),
\label{eq:7}
\end{eqnarray}
 \begin{eqnarray}
	\frac{\delta L}{\delta \rho\left(\omega_{i}\right)}=\gamma p_{0}\beta E\cos\left(\omega\right)-\mu\left(\omega\right)+k_{B}T\ln\left(\rho\left(\omega_{i}\right)\Delta\Omega_{i}/N\right)+V_{w}\lambda =0.
\label{eq:8}
\end{eqnarray}
The first boundary condition is $\psi\left(r\rightarrow\infty\right)=0$, which fixes the origin of the electric potential at $r\rightarrow\infty$.  Other boundary conditions are  $n_{i}\left(r\rightarrow\infty\right)=n_{ib}$ and   $\lambda\left(r\rightarrow\infty\right)=\lambda_{b}$, where  $n_{ib}$  and $\lambda_{b}$  represent the bulk ionic concentration and the Lagrange parameter at  $r\rightarrow\infty$, respectively. 

Using the boundary conditions,  we get the below equations from Eq.(\ref{eq:7}), (\ref{eq:8})
 \begin{eqnarray}
	\frac{n_{i}}{N}=\frac{n_{ib}}{N_{b}}\exp\left(-hV_{i}-e_{0}z_{i}\psi\right),    \left(i=1\cdots m\right).
\label{eq:9}
\end{eqnarray}

\begin{eqnarray}
	\frac{n_{w}}{N}=\frac{n_{wb}}{N_{b}}\exp\left(-h V_{w}\right)\langle\exp\left(-\gamma p_{0}\beta E\cos\left(\omega\right)\right)\rangle_{\omega}.
\label{eq:10}
\end{eqnarray}
where $h\equiv\lambda-\lambda_{b}$ and \cite%
{Iglic_bioelechem_2010, Gongadze_bioelechem_2012}
\begin{eqnarray}
	\left\langle\exp\left(-\gamma p_{0}\beta E\cos\left(\omega\right)\right)\right\rangle_{\omega}=\frac{2 \pi\int^{0}_{\pi}d\left(cos\left(\omega\right)\right)\exp\left(-\gamma p_{0}\beta E\cos\left(\omega\right)\right)}{4\pi}=\frac{\sinh\left(\gamma p_{0} E \beta\right)}{\gamma p_{0} E \beta}.
\label{eq:11}
\end{eqnarray}

Summing  Eqs. (\ref{eq:9}),(\ref{eq:10}) over all species of particles and considering Eqs. (\ref{eq:5}),(\ref{eq:3}) result in the following equations for the number densities of ions and water molecules
\begin{equation}
	n_{i}=\frac{n_{ib}\exp\left(-V_{i}h-e_{0}z_{i}\beta\psi\right)}{\sum^{m}_{k=1} V_{k}n_{kb}\exp\left(-V_{k}h-e_{0}z_{k}\beta\psi\right)+V_{w}n_{wb}\exp\left(-V_{w}h\right)\left\langle\exp\left(-\gamma p_{0}\beta E\cos\left(\omega\right)\right)\right\rangle_{\omega}},    \left(i=1\cdots m\right),
\label{eq:14}
\end{equation}

\begin{equation}
	n_{w}=\frac{n_{wb}\exp\left(-V_{w}h\right)\langle\exp\left(-\gamma p_{0}\beta E\cos\left(\omega\right)\right)\rangle_{\omega}}{\sum^{m}_{k=1} V_{k}n_{kb}\exp\left(-V_{k}h-e_{0}z_{k}\beta\psi\right)+V_{w}n_{wb}\exp\left(-V_{w}h\right)\left\langle\exp\left(-\gamma p_{0}\beta E\cos\left(\omega\right)\right)\right\rangle_{\omega}},
\label{eq:15}
\end{equation}

\begin{eqnarray}
	\sum^{m}_{i=1}n_{ib}\left(\exp\left(-V_{i}h-e_{0}z_{i}\beta\psi\right)-1\right)+n_{wb}\left(\exp\left(-V_{w}h\right)\frac{\sinh\left(\gamma p_{0}\beta E\right)}{\gamma p_{0}\beta E}-1\right)=0.
\label{eq:16}
\end{eqnarray}

In the case when the ions and water molecules have the same size and the surface charge density is negative, that is,  when $V_{+}=V_{-}=V_{w}$ and $\sigma<0$, we can recover all basic equations of \cite%
{Gongadze_bioelechem_2012}.  When we neglect orientational ordering of water dipoles, our approach is identical to that of \cite%
{Boschitsch_jcomchem_2012}. If we assume that $h$  is nearly close to zero, the number densities of ions and water molecules reduce to the corresponding ones of \cite%
{Gongadze_electrochimica_2015}, as shown in eq.(\ref{eq:17}).

\begin{eqnarray}
	n_{i}=\frac{n_{ib}\exp\left(-e_{0}z_{i}\beta\psi\right)}{\sum^{m}_{k=1} V_{k}n_{kb}\exp\left(-e_{0}z_{k}\beta\psi\right)+V_{w}n_{wb}\left\langle\exp\left(-\gamma p_{0}\beta E\cos\left(\omega\right)\right)\right\rangle_{\omega}},    \left(i=1\cdots m\right).
\label{eq:17}
\end{eqnarray}

The Euler$-$Lagrange equation for  $\psi\left(r\right)$  yields the Poisson-Boltzmann equation
\begin{eqnarray}
	\nabla\left(\varepsilon_{0}\varepsilon_{r}\nabla\psi\right)=-e_{0}\sum^{m}_{i=1}z_{i}n_{i},	
\label{eq:18}
\end{eqnarray}
where
 \begin{eqnarray}
	\varepsilon_{r} \equiv n^2+\frac{{\bf P}}{\varepsilon_{0}{\bf E}}.
\label{eq:19}
\end{eqnarray}
 Here, ${\bf P}$ is the polarization vector due to a total orientation of point-like water dipoles. 
From the planar symmetry of this problem, one can see that the electric field strength is perpendicular to the charged surface and have the same magnitude at all points equidistant from the charged surface. The $x$ axis points in the direction of the bulk solution and is perpendicular to the charged surface. Consequently, along the $x$ axis {\bf E} and {\bf P}  have only an $x$ component and {\bf P}  is given as \cite%
{Gongadze_bioelechem_2012}
\begin{eqnarray}
	{\bf P}\left(x\right)=c_{w}\left(x\right)\left(\frac{2+n^2}{3}\right)p_{0}\mathcal{L}\left(\gamma{p_{0}}E\beta \right)\hat{{\bf e}}, 
\label{eq:20}
\end{eqnarray}
where a function $\mathcal{L}\left(u\right)=\coth\left(u\right)-1/u$ is the Langevin function, $\hat{{\bf e}}={{\bf E}/E}$ and $\beta=1/\left(k_{B}T\right)$ .

The electrostatic potential and the number densities of the ions and water molecules are obtained by solving Eqs. (\ref{eq:14}),(\ref{eq:16}),(\ref{eq:18}).

\section{Results and Discussion}
\large
Under the boundary conditions $\psi\left(x\rightarrow\infty\right)=0$  and $E\left(x=0\right)=\sigma/\left(\varepsilon_{0}\varepsilon_{r}\left(x=0\right)\right)$, we combine Eqs. (\ref{eq:14}), (\ref{eq:16}), (\ref{eq:18}) and solve these differential equations for $n_{i}\left(i=1\ldots m\right), n_w, \psi$ by using the fourth order Runge-Kutta method. For all the calculations, the temperature $T$ and the number density $n_{wb}$ of water molecules in the bulk electrolyte have been taken equal to $298.15K$ and  $55mol/l$, respectively. All coions are assigned a charge of $-1e$ and sized to $0.03nm^3$ because previous studies[18,19, 29] showed that the properties of electric double layer are insensitive to size of coions. As in \cite%
{Gongadze_bioelechem_2012, Gongadze_electrochimica_2015}, the water dipole moment should be $3.1D$ so that far away from the charged surface, $x=\infty$, the relative permittivity of the electrolyte reaches the value of pure water.  

\begin{figure}
\begin{center}
\includegraphics[width=0.4\textwidth]{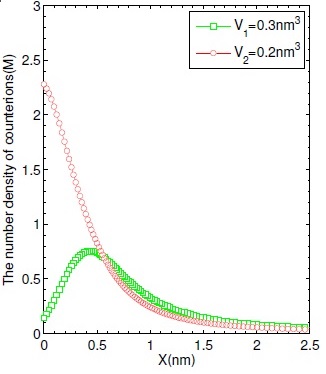}
\caption{\large (Color online) The number densities of counterions as a function of the distance from the charged surface in a mixture of counterions of different species. The bulk concentrations and volumes of counterions are $n_{1b}=6mM, n_{2b}=4mM, V_{1}=0.3nm^3$ and $V_{2}=0.2nm^3$, respectively. The ionic valence of all counterions and surface charge density of the charged surface are  $z_{1}=z_{2}=+1$ and $\sigma=-0.2C/m^2$, respectively.}
\label{fig:1}
\end{center}
\end{figure}

Fig. \ref{fig:1} shows the spatial dependence of the number densities of counterions near a charged surface with surface charge density $\sigma=-0.2C/m^2$. Solid line and Circles stand for the number densities of counterions having $V_{1}=0.3nm^{3}$ and $V_{2}=0.2nm^{3}$ within our approach, respectively. Due to the negative charge density, the negative ions are nearly completely depleted from the region near the charged surface, while the positively charged ions are accumulated in this region.  
Our new approach very well represents a stratification of counterions which was confirmed in experimental and theoretical studies \cite%
{Shapovalov_jphyschem_2006, Shapovalov_jphyschem_2007}, \cite%
 {Chu_biophys_2007}-
\cite%
{Boschitsch_jcomchem_2012}.
Fig. \ref{fig:1} also shows that although bulk concentration of counterions with the larger size is larger than ones of counterions with the smaller size, near the charged surface the number density of the counterions with the larger size is smaller than ones of counterions with the smaller size. 

\begin{figure}
\includegraphics[width=1\textwidth]{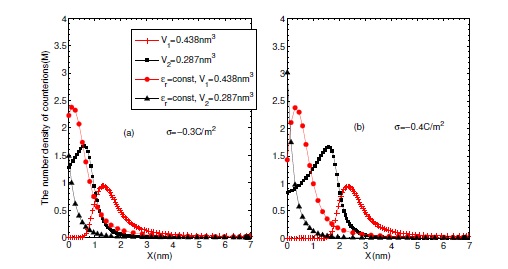}
\caption{\large (Color online) The number densities of two species of counterions as a function of the distance from the charged surface for a surface charge from $-0.3$ and $-0.4C/m^2$. The bulk concentrations and ionic valence of all counterions are $n_{1b}=9mM, n_{2b}=1mM$ and $z_{1}= z_{2}=+1$, respectively.}
\label{fig:2}
\end{figure}

Fig. \ref{fig:2}(a) and Fig. \ref{fig:2}(b) show the number densities of counterions for a mixture of 1 mM CsCl and 9 mM LiCl  for different surface charge densities, considering or not orientational ordering of water dipoles. Plus signs and Squares stand for the number densities of $Li^+$ and $Cs^+$, respectively, for the case when orientational ordering of water dipoles is considered. Circles and Triangles exhibit the number densities of counterions, respectively, for the case when orientational ordering of water dipoles is not considered. To check validity of our approach and compare with experimental and theoretical studies, bulk concentrations of salts are chosen as $n_{1b}=9mM$ and $n_{2b}=1mM$ as in \cite%
{Biesheuvel_jcolloid_2007}, respectively. We considered the volumes of ions by using $V=\left(2\cdot R\right)^{3}$, where $R$ is the hydrated radius of an ion. When the radii of $Li^+$ and $Cs^+$ are chosen as 0.38nm and 0.33nm as in \cite%
{Biesheuvel_jcolloid_2007}, the ionic volumes correspond to $V_{1}=0.438nm^3$ and $V_{2}=0.287nm^3$. However, the authors of \cite%
{Biesheuvel_jcolloid_2007} did not consider orientational ordering of water dipoles.

For both cases studied by our new approach, the stratification of counterions is clearly illustrated.

On one hand, in the case when orientational ordering of water dipoles is not considered, the number densities of counterions are in good agreement with \cite%
{Biesheuvel_jcolloid_2007}, which was obtained by a modified Poisson-Boltzmann theory that empirically includes an extended Carnahan-Starling equation-of-state to account for hard-sphere interactions in electrical double layers containing ions of different size and charge. This fact shows that our approach is reliable and useful.

On the other hand, when orientational ordering of water dipoles is considered, the counterion distributions are expanded into a wider spatial range and the ion concentration peaks are located in farther positions from the charged surface. 
In fact, the competition between orientational ordering of water dipoles and counterion condensation lowers maximum possible ionic concentration \cite%
{Sin_electrochimica_2015, Gongadze_genphysiol_2013, Sin_pccp_2016}. As a result, in order to neutralize the surface charge, more counterions should occupy farther positions.  
In particular, it is noticeable that counterions having the larger volume are nearly depleted from a wide region compared to the case not accounting for orientational ordering of water dipoles.
The above behavior can be explained by the following facts.
Firstly, eq.(\ref{eq:14}) yields an interesting relation for counterions of the same ionic valence,
 \begin{eqnarray}
	\frac{n_{2}}{n_{1}}=\frac{n_{2b}}{n_{1b}}\cdot\exp\left[\left(V_{1}-V_{2}\right)h\right].
\label{eq:21}
\end{eqnarray}

Next, it is well known that accounting for orientational ordering of water dipoles allows electric potential to increase at any position near a charged surface \cite%
{Gongadze_bioelechem_2012, Sin_pccp_2016}.
Finally, the definition of the parameter h gives the fact that as the electric potential approaches zero, the parameter $h\rightarrow0$. In other words, the parameter $h$ increases with increasing the magnitude of the electric potential.
Combining the above three facts, it can be concluded that when orientational ordering of water dipoles is considered, the ratio of the number densities of ions increases and hence counterion stratification is enhanced.

\begin{figure}
\includegraphics[width=1\textwidth]{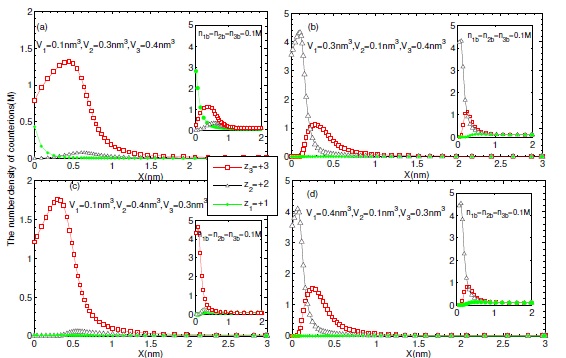}
\caption{\large (Color online) The number densities of counterions as a function of the distance from the charged surface in a mixture of ions with different volumes for surface charge density $\sigma=-0.3C/m^2$. Here bulk concentrations of positive ions are equal to 0.001M, respectively. In the insets, for  bulk concentration of 0.1M the number densities of positive ions are shown, respectively. }
\label{fig:3}
\end{figure}

Recently, the authors of \cite%
{Li_pre_2011, Li_pre_2012} proposed that the ordering of the ion concentration peaks near a surface can be determined by considering the valence to volume ratio, $a=z /V$. However, the authors did consider only case when near a charged surface the relative permittivity of an electrolyte is constant at any position close to the surface. Thus, it is necessary to check the significance of key factors for counterion stratification in the case when the permittivity varies with the distance from the charged surface. Fig. \ref{fig:3} shows the number densities of counterions for the case when non-uniform size effect and orientational ordering of water dipoles are taken into account. Circles, Triangles and Squares correspond to  the number densities of monovalent, divalent and trivalent counterions, respectively. Here the bulk concentration for every species of counterions is $0.001M$. The counterion species are enumerated such that $z_{i} =+i$.  The volumes of the counterions are assigned to produce varying valence to volume ratios.  For the sake of clarity, the following four cases are considered. 

     Case (a):  $ a1 : a2: a3= 1:0.67:0.75$

     Case (b):   $a1 : a2: a3= 0.33:2:0.75$

     Case (c):   $a1 : a2: a3= 1:0.5:1$

     Case (d):   $a1 : a2: a3= 0.25:2:1$

  Thus for case (a) depicted in Fig. \ref{fig:3}(a) the ordering of the ion concentration peaks from the surface corresponds to the $+1e$ species followed by the $+3e$ and then $+2e$ species. This is consistent with the ordering (from large to small) of the ratios listed above. Similarly for cases (b) and (d) the peak ordering from $+2e$ to $+3e$ to $+1e$ predicted by the valence to volume ratios are displayed in Fig. \ref{fig:3}(b),(d). This can be explained by the fact that the competition between entropy, due to translational and orientational arrangements of ions and water molecules, and electrostatic interaction leads to a compromise that in the first layer of stratification, the ionic species with a higher valence-to-volume ratio are contained larger amount than another species. As one can see in Fig.\ref{fig:3}(c), although the ratios for the $+3e$ and $+1e$ species are equal to each other, ions of larger ionic valence predominate in the first layer of electric double layer. In consequence, it can be verified that regardless of whether orientational ordering of water dipoles is considered or not,  the parameter $a$ still plays role of key factor for the ordering of the ion concentration peaks. In Fig. \ref{fig:3}(b),(c),(d), counterion stratification causes $+1e$ species to be depleted from the region close to the charged surface. The results for bulk concentrations of counterions, 0.1M, are in good agreement with corresponding behavior for the bulk concentrations of counterions, 0.001M. 
Although our mean-field approach predicts the counterion stratification,  the results on counterion profiles and concentration peaks (counterion stratification) have rather limited value (especially at large surface charge density and ionic valency value) since  the direct interactions play very important role - leading to many local concentration peaks as shown in \cite%
{Marcovitz_jchemphys_2015}.
\begin{figure}
\begin{center}
\includegraphics[width=1\textwidth]{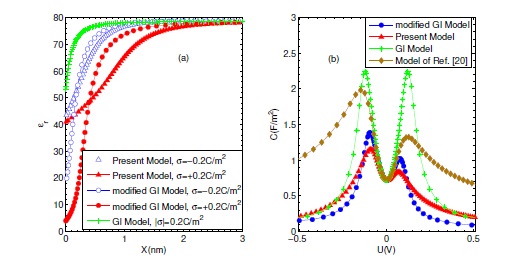}
\caption{\large (Color online) The relative permittivity (a) as a function of the distance from the charged surface and the differential capacitance (b) as a function of surface voltage for our approach($V_{1}= 0.3nm^3, V_{2}= 0.1nm^3$), modified Gongadze-Iglic model($V_{-}= 0.33nm^3, V_{+}= 0.15nm^3$)  and Gongadze-Iglic model($V_{-}= V_{+}= V_{w}$). The ionic valency of ions are $z_{1}=-1e$ and $z_{2}=+1e$. The bulk concentration of binary electrolyte is 0.1M. }
\label{fig:4}
\end{center}
\end{figure}

Fig. \ref{fig:4}(a) displays the relative permittivity near a charged surface with $\sigma= +0.2 C/m^2$ and $\sigma=-0.2 C/m^2$ using our approach, Gongadze-Iglic model \cite%
{Gongadze_bioelechem_2012} and modified Gongadze-Iglic model \cite%
{Gongadze_electrochimica_2015}.  Our model and modified Gongadze-Iglic model explicitly shows asymmetry of the relative permittivity with respect to sign of surface charge. One can observe that near the charged surface the orientation of water molecules and depletion of water molecules result in spatial variation of permittivity \cite%
{Butt_2003, quiroga_jelechemsoc_2014}. 
Fig. \ref{fig:4}(a) displays that within the two models, for $\sigma > 0$  the relative permittivity near the charged surface is smaller than for $\sigma < 0$. As emphasized in \cite%
{Sin_electrochimica_2015}, this fact should be explained by combining the decrease in number density of water molecules and increase in electric field strength. However, near the charged surface there are quite large differences between relative permittivities obtained by two models. First, our model results in slow and small decrease of relative permittivity compared to the modified Gongadze-Iglic model. It is due to the fact that under same circumstance, our model predicts low number density of counterions compared to the modified Gongadze-Iglic model. 
\begin{figure}
\begin{center}
\includegraphics[width=1\textwidth]{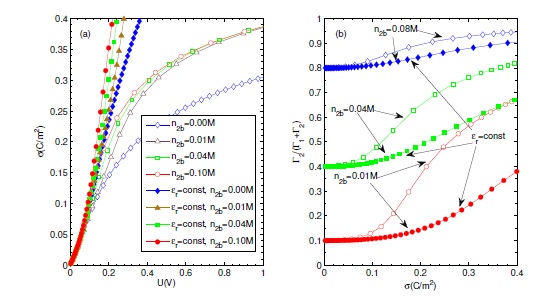}
\caption{\large (Color online) For mixtures having negative ions of different bulk concentrations,  (a) surface charge density as a function of the surface voltage.
(b) the excess portion of negative ions having  volume of $V_{2}= 0.1nm^3$ (b) in electric double layer as a function of the surface charge density when considering orientational ordering of water dipoles or not. For (a) and (b), volumes of negative ions are $V_{1}=0.3nm^3$ and $V_{2}=0.1nm^3$, respectively.  The bulk concentrations for positive ions and negative ions having  volume of $V_{1}$ are equal to 0.1M and $n_{1b}=0.1-n_{2b}$, respectively. The ionic valency of ions are $z_{1}=z_{2}=-1e$ and $z_{3}=+1e$, respectively.}
\label{fig:5}
\end{center}
\end{figure}
In fact, as shown in Eqs. (\ref{eq:14}),(\ref{eq:15}), the term exp$\left(-V_{i}h\right)$ is just a factor lowering number densities of particles.  Since $V_{i}$ is typically larger than $V_{w}$, the value of exp$\left(-V_{i}h\right)$ becomes smaller than exp$\left(-V_{w}h\right)$. In consequence, for the case of our model, the number density of counterions is small compared to one calculated by the modified Gongdze-Iglic model. The fact means that there are more water molecules near the charged surface. Finally, according to Eqs. (\ref{eq:19}), (\ref{eq:20}), the relative permittivity obtained by our model are larger than ones of modified Gongadze-Iglic model. On the other hand, to neutralize the charged surface, more counterions are located in the distant region. Therefore, in the distant region from the charged surface, our model predicts lower number density of water molecules than for the modified Gongadze-Iglic model and hence our model results in low relative permittivity compared to the modified Gongadze-Iglic model. 

\begin{figure}
\begin{center}
\includegraphics[width=1\textwidth]{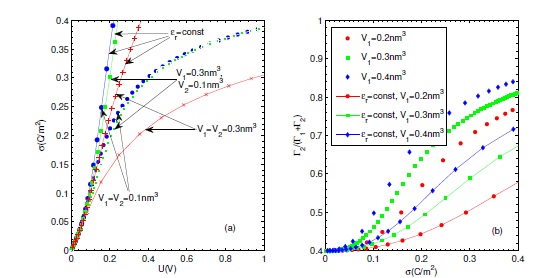}
\caption{\large (Color online) For mixtures of electrolytes having negative ions of different volume, surface charge density (a) as a function of the  surface voltage and excess portion of negative ions having  volume of $V_{2}= 0.1nm^3$ (b) in electric double layer as a function of the surface charge density. The bulk concentrations of negative ions for two species are $n_{1b}=0.06M$ and $n_{2b}=0.04M$, respectively. The bulk concentration of positive ions is 0.1M. The ionic valency of ions are $z_{1}=z_{2}=-1e$ and $z_{3}=+1e$, respectively. }
\label{fig:6}
\end{center}
\end{figure}
Fig. \ref{fig:4}(b) shows the voltage dependences of differential capacitance calculated using our model, Gongadze-Iglic model \cite%
{Gongadze_bioelechem_2012}, modified Gongadze-Iglic model \cite%
{Gongadze_electrochimica_2015} and model of \cite%
{Boschitsch_jcomchem_2012}, respectively.  It is shown that accounting for different sizes of positive and negative ions within the present approach may represent some characteristics of the experimentally observed asymmetric camel-like dependence of differential capacitance on surface voltage \cite%
{Grahame_amchem_1954, damaskin_jsolidstate_2011, foresti_jelecanal_1997}.
As already stated in \cite%
{Sin_electrochimica_2015}, the large size of counterions and the saturated orientational ordering of water dipoles cause the early onset of saturation of counterionic density and lowering of maximum capacitance between negative and positive voltages. A comparison of the curve for our model with one of the modified Gongadze-Iglic model shows that due to the same reason as in Fig. \ref{fig:4}(a), our model predicts slowly varying differential capacitance compared to one for the modified Gongadze-Iglic model.

Fig. \ref{fig:5}(a) shows the surface charge density as a function of the surface voltage for different composition of mixture of electrolytes, considering or not orientational ordering of water dipoles. For the case considering orientational ordering of water dipoles, the surface charge densities are smaller than corresponding ones for the case not considering orientational ordering of water dipoles \cite%
{Gongadze_bioelechem_2012, Sin_pccp_2016}. In the low surface voltage region of U$<$0.lV, surface charge density are nearly equal to each other, irrespective of molar composition of species in mixture. This is due to the fact that in the region, size effects and orientational ordering of water dipoles can be neglected. In the immediate surface voltage region of 0.1V$<$U$<$0.8V, for the cases of multicomponent mixtures($n_{1b}\neq 0$ and $n_{2b}\neq 0$) of electrolytes, surface charge density is in  between corresponding ones for two binary electrolytes($n_{1b}=0$ or $n_{2b}=0$). It can be understood by size effects of counterions. At the high surface voltages of U$>$0.8V, for all of molar compositions of multicomponent mixture, surface charge densities are equal to the corresponding one of binary electrolyte containing counterions with size of $V_{1}=0.1nm^3$. It is attributed to counterion stratification enhanced by orientational ordering of water dipoles. In other words, the stratification of counterions causes more counterions having smaller size to be attracted to the charged surface compared to ones having larger size.   

Fig. \ref{fig:5}(b) displays intuitively the excess portion of counterions having smaller size in electric double layer formed near the charged surface, where the excess charge is defined as follows.
 \begin{eqnarray}
	\Gamma_{i}=\int_{0}^{\infty}\left(n_{i}-n_{ib}\right)dx.
\label{eq:22}
\end{eqnarray}
 For all cases of different molar composition, it is shown that orientational ordering of water dipoles enhances the stratification of counterions. In particular, Fig. \ref{fig:5}(b) shows that for the case of low bulk concentration of the counterion having small size($V_{2}$), an increase in the excess charge is larger than corresponding ones for high bulk concentration of the counterion.  A comparison Fig. \ref{fig:5}(a) with Fig. \ref{fig:5}(b) says that although at high surface voltages surface charge densities are equal to each other, molar compositions strongly differ from each other.

Fig. \ref{fig:6}(a) shows the surface charge density according to the surface voltage for the cases of mixture of electrolytes having counterions with different sizes. It is noted that the values for $V_{1}=0.3nm^3, V_{2}=0.1nm^3$ are close to ones for $V_{1}=0.1nm^3, V_{2}=0.1nm^3$ and largely differ from ones for  $V_{1}=0.3nm^3, V_{2}=0.3nm^3$, regardless of considering orientational ordering of water dipoles or not. In particular, for the case considering orientational ordering of water dipoles, such behaviors are represented more clearly.  As in Fig. \ref{fig:5}(a), the fact is due to the stratification of counterions enhanced by orientational ordering of  water dipoles. 

Fig. \ref{fig:6}(b) shows the excess portion of counterions of smaller size($V_{2}$) in electric double layer formed near the charged surface. For different size of large counterion($V_{1}$), it is shown that orientational ordering of water dipoles enhances the stratification of counterions.  It also represents that the larger difference $V_{1}-V_{2}$ in sizes of counterions yields the larger excess portion of smaller ions in the electric double layer. The fact can be explained by the same arguments as previously stated in Fig. \ref{fig:2}(a),(b).
\begin{figure}
\begin{center}
\includegraphics[width=1\textwidth]{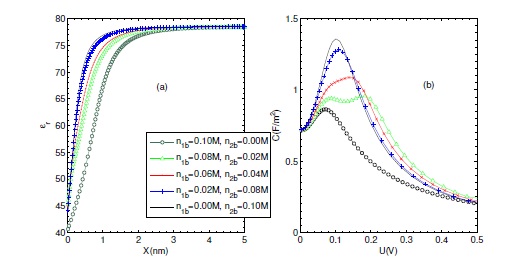}
\caption{\large (Color online) The relative permittivity (a) as a function of the distance from the charged surface and differential capacitance (b) as a function of surface voltage for our approach. Other parameters are the same as in Fig. \ref{fig:5}.}
\label{fig:7}
\end{center}
\end{figure}

\begin{figure}
\begin{center}
\includegraphics[width=1\textwidth]{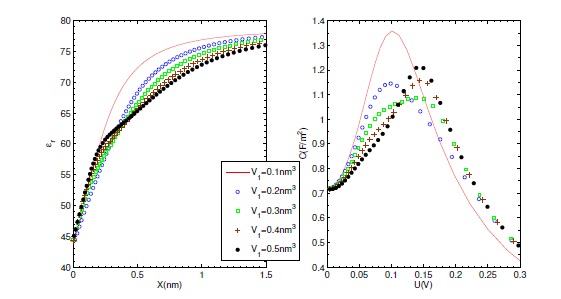}
\caption{\large (Color online) The relative permittivity (a) as a function of the distance from the charged surface and differential capacitance (b) as a function of surface voltage for our approach. Other parameters are the same as in Fig. \ref{fig:6}.}
\label{fig:8}
\end{center}
\end{figure}

Fig. \ref{fig:7}(a) shows the surface charge density as a function of the distance from the charged surface for different molar composition of mixture of electrolytes when considering orientational ordering of water dipoles. Although the bulk concentration of counterions having smaller size is low compared to ones of counterions with larger size($n_{1b}=0.08M, n_{2b}=0.02M$), the relative permittivity at charged surface is close to one for the case ($n_{1b}=0.00M, n_{2b}=0.10M$) rather than for the case of ($n_{1b}=0.00M, n_{2b}=0.10M$). This is understood from the fact that due to $V_{1}>V_{2}$ and Eq.(\ref{eq:21}) , the population of counterions with volume of $V_{2}$ near the charged surface is larger than for counterions having volume of $V_{1}$.  

Fig. \ref{fig:7}(b) shows the differential capacitance as a function of the positive surface voltage for different molar composition of mixture of electrolytes when considering orientational ordering of water dipoles. Here two facts are of interest.  First, in the voltage region  of 0.2$<$U, for multicomponent electrolytes, the differential capacitances are higher than ones for binary electrolytes. Secondly, it is noticeable that for the case of ($n_{1b}=0.08M, n_{2b}=0.02M$), the differential capacitance curve contains two peaks. Furthermore, it is interesting that for case of ($n_{1b}=0.08M, n_{2b}=0.02M$), differential capacitance is kept nearly constant in relatively wide region of surface voltage. It should be explained by counterion stratification. As shown in Fig. \ref{fig:5}(b), when the bulk concentration of counterions with small size($V_{2}$) becomes lower, counterion stratification becomes stronger. In other words, a stronger stratification of counterions yields a larger change in composition of mixiture.  As a result, a stronger stratification of counterions makes charging process of the charged surface easier and hence results in a larger capacitance. In addition, this can cause the fact that although $d\sigma/dU>0$ is always satisfied, in two regimes $d^{2}\sigma/d^{2}U<0$.  In consequence, the above facts are understood by the stratification of counterions.

Fig. \ref{fig:8}(a) shows the surface charge density as a function of the distance from the charged surface for different size of counterion in mixture of electrolytes when considering orientational ordering of water dipoles.
In the closest region($x<0.4nm$) near the charged surface, for mixture having counterions of larger size, the relative permittivity is large compared to one for mixtures containing counterions with smaller size.
In farther region, the order for the relative permittivities is reversed. This can be explained by the role of the term exp$\left(-V_{i}h\right)$ as above mentioned.  First, Eq.(\ref{eq:21}) says that when $V_{1}$ becomes much larger, more counterions having $V_{2}$ are located near the charged surface  compared to the original case. Accounting for $V_{1}>V_{2}$, this means that there are more water molecules near the charged surface and hence the relative permittivity increases. In farther region, counterions having size of $V_{1}$ exist relatively more than ones for closest region. In consequence, the number density of water molecules is decreased and hence the relative permittivity decreases. 

Fig. \ref{fig:8}(b) shows the differential capacitance as a function of the positive surface voltage for different size of counterions in mixture of electrolytes when considering orientational ordering of water dipoles.
In the low voltage region of $U<0.15V$, for mixture having counterions of larger size, the differential capacitance is small compared to one for mixtures containing counterions of smaller size. This is attributed to the fact that counterions having larger size cause surface voltage for a surface charge density to become high compared to ones with smaller size.
In the region of $U>0.15V$, the order for the differential capacitances is reversed. In fact, counterion stratification makes percentage of counterions with smaller size in electric double layer to increase. An increase in number density of counterions with small size yields the fact that a small increase of surface voltage causes relatively high increase in surface charge density compared to the original case. In consequence, for the case of larger size of counterions the differential capacitance is higher than for the smaller size.

To further amend our model, it is necessary to include also the Stern layer accounting for unequal distances of closest approach for positive and negative ions \cite%
{iglic_intjelectrol_2015, yu_jcolloid_2006, Outhwaite_jchemphysl_1986}. Additionally, although our model is a mean field approach, it will be also interesting to account for direct interaction for multicomponent electrolytes.  This amendment would change both the relative height and voltages for both maxima of the differential capacitance camel-like curve.

\section{Conclusions}
In this work, the non-uniform size effects of ions and water molecules in the electric double layer are described by modifying previously published mean-field models of \cite%
{Gongadze_bioelechem_2012} which took into account orientational ordering of water dipoles in saturation regime near the charged surface and asymmetry in ion size in binary electrolyte solution within lattice statistics approach  \cite%
{Gongadze_electrochimica_2015}. Similarly as in  \cite%
{Gongadze_electrochimica_2015}, we have partially abandoned our previous assumption of small volume shares of ions  \cite%
{Sin_electrochimica_2015} everywhere in the electrolyte solution.

In contrast to previously modified GI model \cite%
{Gongadze_electrochimica_2015}, within our modified GI model, for both cases when orientational ordering of water dipoles is considered or not, the stratification of counterions for mixtures of counterions of different species has been indicated and then the significance of key factors of counterion stratification has been checked and recognized. Furthermore, it is predicted that orientational ordering of water dipoles can enhance counterion stratification. 

The influence of size asymmetry of negative and positive ions in electrolyte solution on relative permittivity and on differential capacitance of EDL in binary electrolyte solution has been also investigated and compared to the previous results  \cite%
{Gongadze_bioelechem_2012, Gongadze_electrochimica_2015}. 

The effect of the difference in sizes of counterions and  different bulk concentrations of counterions on surface charge density, excess portion of counterions, relative permittivity and differential capacitance are very important and can be well understood by using the concept of counterion stratification. 

In practice, our modified approach and results can be used to describe the frontiers of biology, medicine, colloidal science and electrochemistry such as the binding of charged ligands to the membrane surface 
\cite%
{Cevc_biochim_1990, Safran_book_1994}, the interactions of vesicles with the membrane \cite%
{McLaughlin_annrev_1989}, osteoblast attachment to biomaterials \cite%
{Smith_biomed_2004}, RNA folding \cite%
{Draper_annrev_2005} and differential capacitance \cite%
{Wang_physchem_2009,Imani_nanoscale_2015} of electric double layer capacitor for which non-uniform size effects of ions and orientational ordering of water dipoles may be important.  In particular, this approach can be suitable for studying multicomponent mixtures of electrolytes having different species of counterions.


\section{\bf Reference}

\begin{thebibliography}{99}
\large 
\bibitem{Helmholtz_annphys_1879}
H. Helmholtz, Studien uber electrische grenzschichten. Ann. Phys. Chem. {\bf 7} (1879)  337-382.

\bibitem{Gouy_physfrance_1910}
M.G. Gouy , Sur la constitution de la charge \'electrique a la surface d'un \'electrolyte, J. Phys.(France) {\bf 9} (1910)  457-468.

\bibitem{Chapman_philos_1913}
D.L. Chapman,  A contribution to the theory of electrocapillarity, Philos. Mag. {\bf 25} (1913) 475-481.

\bibitem{Shapovalov_jphyschem_2006}
V.L. Shapovalov, G. Brezesinski, Breakdown of the Gouy-Chapman model for highly charged Langmuir monolayers: counterion size effect, J. Phys. Chem. B {\bf 110} (2006) 10032-10040.

\bibitem{Shapovalov_jphyschem_2007}
V.L. Shapovalov, M.E. Ryskin, O.V. Konovalov, A. Hermelink, G. Brezesinski, Elemental analysis within the electrical double layer using total reflection X-ray fluorescence technique, J. Phys. Chem. B {\bf 111} (2007) 3927-3934.

\bibitem{Stern_zelektrochem_1924}
O. Stern, Zur theorie der electrolytischen doppelschicht. Z. Elektrochem. {\bf 30} (1924) 508-516.

\bibitem{Bikerman_philmag_1942}
J.J. Bikerman, Structure and capacity of electrical double layer, Philos. Mag. {\bf 33} (1942) 384-397.

\bibitem{Borukhov_prl_1997}
I. Borukhov, D. Andelman, H. Orland,  Steric effects in electrolytes: a modified Poisson-Boltzmann equation, Phys. Rev. Lett. {\bf 79} (1997) 435-438.

\bibitem{Borukhov_electrochimica_2000}
I. Borukhov, D. Andelman,  H. Orland, Adsorption of large ions from an electrolyte solution: a modified Poisson-Boltzmann equation, Electrochim. Acta {\bf 46} (2000) 221-229.

\bibitem{Iglic_physfran_1996}
 V. Kralj-Igli\v c, A. Igli\v c, A simple statistical mechanical approach to the free energy of the electric double layer including the excluded volume effect, J. Phys.France {\bf 6} (1996) 477-491. 

\bibitem{Bohnic_electrochimica_2001}
K. Bohnic, V. Kralj-Igli\v c, A. Igli\v c, Thickness of electrical double layer. Effect of ion size, Electrochim. Acta {\bf 46} (2001) 3033-3040.

\bibitem{Bohnic_bioelechem_2002}
K. Bohnic, A. Igli\v c,  T. Slivnik, V. Kralj-Igli\v c, Charged cylindrical surfaces: effect of finite ion size, Bioelectrochemistry {\bf 57} (2002) 73-81.

\bibitem{Bohnic_bioelechem_2005}
K. Bohnic, J. Gimsa, V. Kralj-Igli\v c, T. Slivnik, A. Igli\v c, Excluded volume driven counterion condensation inside nanotubes in a concave electrical double layer model, Bioelectrochemistry {\bf 67} (2005)  91-99.

\bibitem{Chu_biophys_2007}
V.B. Chu, Y. Bai, J. Lipfert, D. Herschlag, S. Doniach. Evaluation of ion binding to DNA duplexes using a size-modified Poisson-Boltzmann theory, Biophys. J. {\bf 93}  (2007)  3202-3209.

\bibitem{Tresset_pre_2008}
G. Tresset, Generalized Poisson-Fermi formalism for investigating size correlation effects with multiple ions. Phys. Rev. E. 78 ( 2008) 061506 

\bibitem{Kornyshev_physchem_2007}
A.A. Kornyshev, Double-layer in ionic liquids: paradigm change?, J. Phys. Chem. B {\bf 111} (2007) 5545-5557.

\bibitem{Popovic_pre_2013}
M. Popovic, A. Siber, Lattice-gas Poisson-Boltzmann approach for sterically asymmetric electrolytes, Phys. Rev. E {\bf 88} (2013) 022302:1-5.


\bibitem{Li_pre_2011}
S. Zhou, Z. Wang, and B. Li, Mean-field description of ionic size effects with non-uniform ionic sizes: a numerical approach, Phys. Rev. E. {\bf 84} (2011) 021901.

\bibitem{Li_pre_2012}
J. Wen, S. Zhou, Z. Xu, and B. Li, Competitive adsorption and ordered packing of counterions near highly charged surfaces: from mean-field theory to Monte Carlo simulations,  Phys. Rev. E. {\bf 85} (2012) 041406.


\bibitem{Boschitsch_jcomchem_2012}
A.H. Boschitsch,  P.V. Danilov, Formulation of a new and simple non-uniform size-modified Poisson-Boltzmann description, J. Comput. Chem. {\bf 33} (2012) 1152-1164.


\bibitem{Onsager_amchem_1936}
L. Onsager, Electric moments of molecules in Liquids, J. Am. Chem. Soc. {\bf 58} (1936) 1486-1493. 

\bibitem{Kirkwood_chemphys_1939}
J.G. Kirkwood, The dielectric polarization of polar liquids, J. Chem. Phys. {\bf 7} (1939) 911-919.

\bibitem{Booth_chemphys_1951}
F. Booth, The dielectric constant of water and the saturation effect, J. Chem. Phys. {\bf 19} (1951)  391-394.

\bibitem{Booth_chemphys_1955}
F. Booth, Dielectric constant of polar liquids at high field strengths, J. Chem. Phys. {\bf 23} (1955) 453-457. 

\bibitem{Iglic_bioelechem_2010}
A. Igli\v c, E. Gongadze,K. Bohnic, Excluded volume effect and orientational ordering near charged surface in solution of ions and Langevin dipoles, Bioelectrochemistry {\bf 79} (2010) 223-227.

\bibitem{Gongadze_bioelechem_2012}
E. Gongadze, A. Igli\v c, Decrease of permittivity of an electrolyte solution near a charged surface due to saturation and excluded volume effects, Bioelectrochemistry {\bf 87} (2012) 199-203. 

\bibitem{Gongadze_electrochimica_2013}
E. Gongadze, A. Velikonja, T. Slivnik, V. Kralj-Igli\v c, A. Igli\v c, The quadrupole moment of water molecules and the permittivity of water near a charged surface, Electrochim. Acta {\bf 109} (2013) 656-662.

\bibitem{Gongadze_electrochimica_2015}
E. Gongadze, A. Igli\v c, Asymmetric size of ions and orientational ordering of water dipoles in electric double layer model - an analytical mean-field approach, Electrochim. Acta {\bf 178} (2015) 541-545.

\bibitem{Sin_electrochimica_2015}
J.-S. Sin, S.-J. Im, K.-I. Kim, Asymmetric electrostatic properties of an electric double layer: a generalized Poisson-Boltzmann approach taking into account non-uniform size effects and water polarization, Electrochim. Acta {\bf 153} (2015) 531-539.

\bibitem{Biesheuvel_jcolloid_2007}
P.M. Biesheuvel, M. van Soestbergen, Counterion volume effects in mixed electrical double layers, J. Colloid Interface Sci. {\bf 316} (2007) 490-499.

\bibitem{Gongadze_genphysiol_2013}
E. Gongadze, A. Igli\v c, Excluded volume effect of counterions and water dipoles near a highly charged surface to a rotationally averaged Boltzmann factor for water dipoles, Gen. Physiol. Biophys. {\bf 32} (2013) 143-145.

\bibitem{Sin_pccp_2016}
J.-S. Sin, H.-C. Pak, K.-I. Kim, K.-C. Ri, D.-Y. Ju, N.-H. Kim, C.-S. Sin, An electric double layer of colloidal particles in
salt-free concentrated suspensions including non-uniform size effects and orientational ordering of water dipoles, Phys. Chem. Chem. Phys.  {\bf 18} (2016) 234-243.

\bibitem{Butt_2003}
H.J. Butt, K. Graf, M. Kappl, Physics and chemistry of interfaces. 1st ed. (Wiley-VCH Verlag, Weinhelm, 2003).

\bibitem{quiroga_jelechemsoc_2014}
M.A. Quiroga, K.H.  Xue, T.K. Nguyen, M. Tulodziecki, H. Huang, A.A. Franco, A multiscale model of electrochemical double layers in energy conversion and storage devices, J. Electrochem. Soc. {\bf 161} (2014) 3302-3310.

\bibitem{Grahame_amchem_1954}
D.C. Grahame, Differential capacity of mercury in aqueous sodium fluoride solutions: I. Effect of concentration at $25^{\circ}$C,  J. Am. Chem. Soc. {\bf 76} (1954) 4819-4823.

\bibitem{damaskin_jsolidstate_2011}
B.B. Damaskin, O.A. Petrili, Histroical development of theories of the electrochemical double layer, J. Solid. State. Electrochem. {\bf 15} (2011) 1317-1334.

\bibitem{foresti_jelecanal_1997}
M.L. Foresti, M. Innocenti, R. Guidelli, A. Hamelin, Electrochemical investigation of 1, 5-pentanediol adsorption on the Ag(111) and Ag(110) faces, J. Electroanal. Chem. {\bf 467} (1997) 217-229.

\bibitem{iglic_intjelectrol_2015}
A. Velikonja, V. Kralj-Igli\v c, A. Igli\v c, On asymmetric shape of electric double layer capacitance curve, Int. J. Electrochem. Sci. {\bf 10} (2015) 1-7.

\bibitem{yu_jcolloid_2006}
J. Yu, G. Aguilar-Pineda, A. Antillon, S. Dong, M. Lozada-Cassou, The effects of unequal ionic sizes for an electrolyte in a charged slit, J. Colloid Interface Sci. {\bf 295} (2006) 124-134.

\bibitem{Outhwaite_jchemphysl_1986}
C. Outhwaite, L. Bhuiyan, A modified Poisson-Boltzmann equation in electric double layer theory for a primitive model electrolyte with size asymmetric ionsize, J. Chem. Phys. {\bf 84} (1986) 3461-3471. 

\bibitem{Cevc_biochim_1990}
G. Cevc, Membrane electrostatics, Biochim. Biophys. Acta {\bf 1031} (1990) 311-382.

\bibitem{Safran_book_1994}
A. Safran, Statistical Thermodynamics of Surfaces, Interfaces, and Membranes,Addison-Wesley Publishing Company, 1994.

\bibitem{McLaughlin_annrev_1989}
S. McLaughlin, The electrostatic properties of membranes,  Ann. Rev. Biophys. Chem. {\bf 18} (1989) 113-136.

\bibitem{Smith_biomed_2004}
LO. Smith, M.J. Baumann, L.R. McCabe, Electrostatic interactions as a predictor for osteoblast attachment to biomaterials, J. Biomed. Mat. Res. A. {\bf 70} (2004) 436-441.

\bibitem{Draper_annrev_2005}
D.E. Draper, D. Grilley , A.M. Soto,  Ions and RNA folding,  Ann. Rev. Biophys. Biomol. Struct. {\bf 34} (2005) 221-243.

\bibitem{Wang_physchem_2009}
H. Wang, L. Pilon, Accurate simulations of electric double layer capacitance of ultramicroelectrodes, J. Phys. Chem. C {\bf 115} (2011) 16711-16719.

\bibitem{Imani_nanoscale_2015}
 R. Imani, M. Pazoki, A. Tiwari, G. Boschloo, A. Turner, V. Kralj-Igli\v c, A. Igli\v c, Band edge engineering of TiO2@DNA nanohybrids and implications for capacitive energy storage devices, Nanoscale {\bf 7} (2015) 10438-10448.

\bibitem{Marcovitz_jchemphys_2015}
A. Marcovitz, A. Naftaly, Y. Levy,  Water organization between oppositely charged surfaces: Implications for protein sliding along DNA, J. Chem. Phys. {142} (2015) 085102:1 - 10 .

\bibitem{Wicke_zelechem_1952}
 E. Wicke, M. Eigen,  Uber den  Einfluss  des  Raumbedarfs  von  Ionen  in wassriger  Losung  auf   ihre   Verteilung   im elektrischen  Feld und  ihre Aktivitatskoeffizienten,  Z. Elektrochem. 56 (1952) 551-561.
\end{thebibliography}
\end{document}